\documentclass[conference]{IEEEtran}

\makeatletter

\def\ps@IEEEtitlepagestyle{%
  \def\@oddfoot{\mycopyrightnotice}%
  \def\@evenfoot{}%
}
\def\mycopyrightnotice{%
  {\footnotesize XXX-X-XXXX-XXXX-X/XX/\$XX.00~\copyright~20XX IEEE\hfill}
  \gdef\mycopyrightnotice{}
}

\usepackage{blindtext}
\usepackage{eso-pic}
\usepackage{hyperref}
\usepackage{csquotes}
\IEEEoverridecommandlockouts
\usepackage{cite}
\usepackage{amsmath,amssymb,amsfonts}
\usepackage{algpseudocode}
\usepackage{graphicx}
\usepackage{textcomp}
\usepackage{xcolor}
\usepackage{booktabs}
\usepackage{tikz}
\def\BibTeX{{\rm B\kern-.05em{\sc i\kern-.025em b}\kern-.08em
    T\kern-.1667em\lower.7ex\hbox{E}\kern-.125emX}}
    
\usepackage{eso-pic}
\newcommand\AtPageUpperMyright[1]{\AtPageUpperLeft{%
 \put(\LenToUnit{0.17\paperwidth},\LenToUnit{-2cm}){%
     \parbox{0.9\textwidth}{\raggedleft\fontsize{8}{11}\selectfont #1}}%
 }}%
\newcommand{\conf}[1]{%
\AddToShipoutPictureBG*{%
\AtPageUpperMyright{#1}
}
}

\begin{document}
\title{\vspace*{1cm} 
Information Flow Paths from RTL Traces\\\thanks{This work was supported by NSF Award 171858.}}

\author{\IEEEauthorblockN{Calvin Deutschbein}
\IEEEauthorblockA{\textit{School of Computing and Information Sciences} \\
\textit{Willamette University}\\
Salem, Oregon, U.S. \\
ckdeutschbein@willamette.edu}
\and
\IEEEauthorblockN{Owyn Wyatt
}
\IEEEauthorblockA{\textit{School of Computing and Information Sciences} \\
\textit{Willamette University}\\
Salem, Oregon, U.S. \\
orwyatt@willamette.edu}
}

\maketitle
\conf{\textit{ 6. Interdisciplinary Conference on Electrics and Computer (INTCEC 2026) \\ 
24-25 September 2026, Chicago-USA}}
\begin{abstract}

Security validation is an important yet challenging part of the hard-
ware design process, yet by convention validation engineers are
tasked with defining the threat model, specifying
the relevant security properties, detecting any violations of those
properties, and assessing the consequences to system security, each of
which is manually intensive and may introduce errors. The combined technologies
of information flow tracking and specification mining represent an automated
approach to property generation and validation, but prior work on 
information flow tracking on RTL trace data~\cite{deutschbein21} was limited
to find cases under which information flowed between registers, without
reproducing full paths to capture how sensitive information propagates through
a design. With the introduction of new technologies accelerating hardware
analysis~\cite{deutschbein25hwcicd, deutschbein2025vcd2df},
we develop a novel approach for constructing information flow paths
from register transfer level (RTL) trace data.
\end{abstract}


\section{Introduction}

The discovery of transient execution CPU vulnerabilities in 2018~\cite{kocher18, lipp18}
demonstrated the need for novel hardware validation frameworks -  ones with sufficient
expressability to capture hyperproperties~\cite{clarkson2008}. 
The first two discovered vulnerabilites, Spectre and Meltdown, 
both allowed disclosure of sensitive
information at the level of microarchitecture and timing channels. 
To capture properties of this type, it is necessary to specify
behavior across multiple traces, or multiple runs of execution, requiring
more advanced validation technologies.


Information flow tracking (IFT)~\cite{ardeshiricham17,becker2017,hu2014,hu2016,hu2018}
offers a promising solution for validation. Essentially, every signal (register or wire)
in a design is granted a corresponding ``shadow" version which, rather than storing
data, stores a tracking status. Any signals containing sensitive data
may have their tracking status set to nonzero, then a device may be simulated and the resultant tracking status may be found for each
signal.

However, this tracking logic does not differentiate immediate and 
intermediate sources of information. If a key is leaked into 
an output register, it is likely via intermediate registers that eluded analysis.
Information flow tracking excels at determining whether information leaks, but not the path.

In this work, we
analyze information flow traces - traces of execution which are generated by executing, in simulation,
not only a hardware design but also additional state that instruments the design with information flow tracking. Using simulated traces at register transfer level (RTL), we capture
micro-architectural information flow paths. We will:

\begin{enumerate}
    \item Describe our problem statement.
    \item Describe our proposed solution algorithmically.
    \item Outline our implementation of the proposed solution.
    \item Examine the PicoRV32 RISC-V CPU design.
\end{enumerate}

\section{Problem Statement}

In this work, we will address what we term the path reconstruction problem for information flow.

\textbf{Path Reconstruction Problem}: Given a track of execution of a hardware design that is instrumented with information flow tracking, determine the path of information flow from a given \textit{source signal} to a given \textit{sink signal}.

\begin{figure*}[t]
  \centering
  \includegraphics[width=\textwidth]{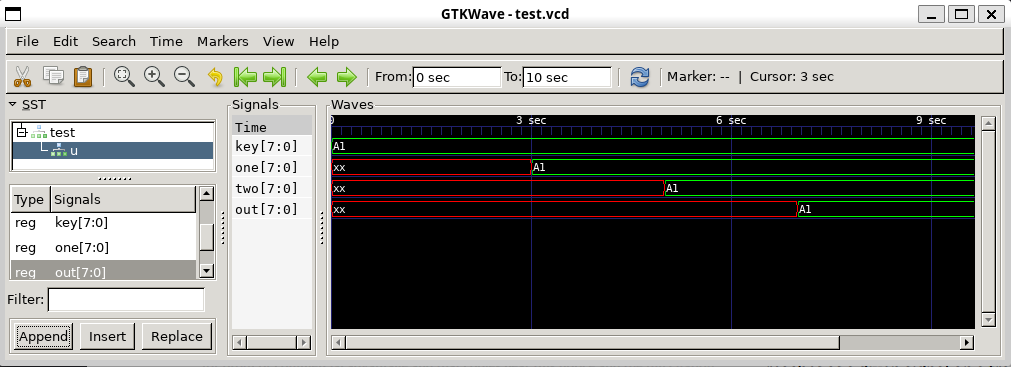}
  \caption{The waveform of a design illustrating an information flow from \texttt{key} to \texttt{out} for which the path is ambiguous.}
  \label{fig:vcd}
\end{figure*}

\subsection{Example}

As a motivating example, a simple design comprised of four registers, which we will consider manually:

\begin{enumerate}
    \item \texttt{key} - A register we think of as holding a secret, such as a private key.
    \item \texttt{one} - An internal register in a module.
    \item \texttt{two} - An internal register in a module.
    \item \texttt{out} - An output potentially visible to unprivileged user.
\end{enumerate}

We take the security goal to be ensuring the contents of \texttt{key} - which we can regard as its literal value, but more properly should be any information about \texttt{key} - never reach \texttt{out}. 

Suppose information about \texttt{key} leaks through \texttt{out}. Suppose in static analysis we find the design contains no cases in which \texttt{key}'s value directly impacts the resultant value of \texttt{out}. Information must pass through some \textit{intermediate signal}.

In our example, \texttt{key} is the \textit{source signal} and \texttt{out} is the \textit{sink signal} while \texttt{one} and \texttt{two} are examples of possible \textit{intermediate signals}. We wish to determine, from simulating the design and evaluating its trace data, whether one or both of these two registers constitute the pathway through which privileged information becomes visible to an unprivileged user.

Consider Figure~\ref{fig:vcd}. Here \texttt{key} has some initial value which propagates to \texttt{one}, \texttt{two}, and \texttt{out}. However, this temporal sequencing is insufficient to determine which pairwise flows (by which we mean an atomic flow from a source to sink at a certain time point) occur at any given time. For example, is \texttt{two} updated with the value of \texttt{key}, with the value of \texttt{one}, or does it coincidentally take on the same value?

Consider instead a design, rather than a trace:

\begin{verbatim}
  initial begin
    key = 8'hA1; // Secret
    cnt = 0;
  end

  always @(posedge clk) begin
    cnt <= cnt + 1;
    if (cnt == 1) 
      one <= key;
    else if (cnt == 2)
      two <= key;
    else if (cnt == 3)
      out <= two;
  end
\end{verbatim}

The information leak occurs when the counter \texttt{cnt} reaches 3 and the value of \texttt{two} is written directly to output. However, we note from the perspective of a single trace, the waveform would be identical if rather than:
\begin{verbatim}
    else if (cnt == 3)
      out <= two;
\end{verbatim}
We instead had a design that contained the vulnerability:
\begin{verbatim}
    else if (cnt == 3)
      out <= one;
\end{verbatim}
In either case we may address matters differently. For example, it may be the case that \texttt{one} is intended to be a valid storage signal for the key value, and that \texttt{two} is an intended staging signal for eventual output, so these are errors of different kinds. 

While a contrived example, rather than working on individual signals, the same occurs over modules rather than signals. Or rather than leakage occurring through assignment, it occurs through branching control flow. 

\section{Methodology}

To determine information flow paths, we propose to:

\begin{itemize}
    \item Instrument a design with information flow tracking.
    \item Map over each \textit{signal} in the design.
    \begin{itemize}
        \item Fix said signal as the \textit{source signal}
        \item Trace the design over some fixed inputs.
        \item Extract \textit{times-of-flow} for all \textit{sink signals} in the trace.
    \end{itemize}
    \item Reduce the \textit{times-of-flow} in to a single dataframe.
    \item Query the dataframe to identify information flow paths.
\end{itemize}

We specifically adapt the notion of the \textit{time-of-flow} from the Isadora specification miner~\cite{deutschbein21,deutschbein2022,deutschbein2023}, which also inspired our usage of information flow tracking. We will define \textit{time-of-flow} rigorously in \autoref{sec:tof} after first introducing information flow tracking.

\subsection{Information Flow Tracking}

Information Flow Tracking (IFT) is a powerful security verification technique that monitors how information moves through a hardware design. IFT has been demonstrated at the RTL and gate level and has been used to monitor implicit flows through digital side channels. For our example, we add simple manual IFT, but our method is agnostic to IFT implementation.

\subsubsection{Tracking Signals}

\begin{figure*}[t]
  \centering
  \includegraphics[width=\textwidth]{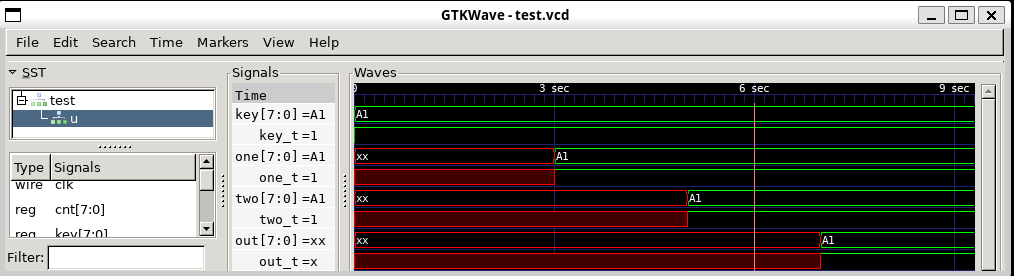}
  \caption{A waveform showing the IFT tracking signals after fixing \texttt{key} as the source.}
  \label{fig:ift}
\end{figure*}

For each signal within the design, we add a corresponding \textit{tracking signal}. We adopt the shorthand of the `\texttt{\_t}` suffix denoting a tracking signal. We use signals of a single bit, to simply denote that information has, or has not, reached the tracked signal.

As a simplifying assumption, we will consider only the four enumerated signals, as follows:

\begin{verbatim}
  output reg [7:0] out;
  reg              out_t;
  reg        [7:0] key;
  reg              key_t;
  reg        [7:0] one;
  reg              one_t;
  reg        [7:0] two;
  reg              two_t;
\end{verbatim}

Then, after any line in which one of these signals is updated based on the value of another, we propagate the tracking label from the \textit{source signal} to the \textit{sink signal}.

\begin{verbatim}
    if (cnt == 1) 
      begin
        one <= key;
        one_t <= key_t | one_t;
      end
    else if (cnt == 2)
      begin
        two <= key;
        two_t <= key_t | two_t;
      end
    else if (cnt == 3)
      begin
        out <= two;
        out_t <= two_t | out_t;
      end
\end{verbatim}

We can imagine less trivial cases:

\begin{verbatim}
      two   <= key - one;
      two_t <= key_t | one_t | two_t;
\end{verbatim}




We provide an illustration in Figure~\ref{fig:ift} when we set the initial label on \texttt{key} to be non-zero.

\subsection{Simulation}

\subsubsection{Testbench}

The mere description of a hardware design cannot be executed; rather it describes the capacity to perform computation over provided inputs. Therefore, in isolation, a hardware design cannot generate a trace of execution. So, we introduce the additional requirement of a testbench. 
\subsubsection{Setting the Source}

Simulating with IFT requires additionally fixing a \textit{root source signal} or signals. To do so, we make a single modification to the module containing the \textit{root source signal}, where it is sufficient to set any given signal or even set of signals to nonzero in the \texttt{initial begin} block, and the design will track information flow. In a security context, we regard this \textit{root source signal} as a designation of sensitive data, and the information flow tracking will then track which signals in the design have values somehow dependent on the value of this \textit{root source signal}.

For instance, having identified \texttt{key} as the appropriate signal to fix as the root, we update the counter as follows:
\begin{verbatim}
  initial begin
    key   = 8'hA1;
    key_t = 1;     // This line is new.
    cnt   = 0;
  end
\end{verbatim}

\subsubsection{Mapping Stage}

To gather sufficient data about the design to construct information flow paths we simulate, and therefore derive trace data with tracking information, for every signal within the design. This computationally costly for larger designs, but is correspondingly more useful on such designs, and may be both automated and parallelized. Ongoing work to generate hardware traces on cloud infrastructure~\cite{deutschbein25hwcicd} was a core enabling technology to drive this insight.

\subsection{Graph Construction}

\begin{figure*}[t]
  \centering
  \includegraphics[width=\textwidth]{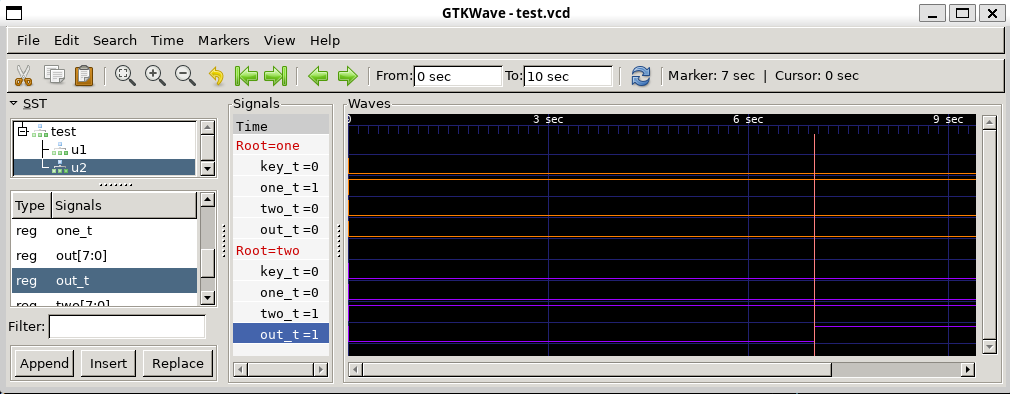}
  \caption{A design showing tracking signals given a \textit{root source signal} of \texttt{one}, in the upper 5 rows in orange, or \texttt{two}, in the lower 5 rows in violet. There is precisely one case of information flow, from \texttt{two} to \texttt{out} at time $\Delta t = 7$ (shown with the horizontal mark) which corresponds to the only \textit{time-of-flow} relative to either of these \textit{root source signals.}}
  \label{fig:all}
\end{figure*}

Given the trace data, we can identify \textit{times-of-flow} and organize these as graph edges.

\subsubsection{\textit{Times-of-flow} Definition}
\label{sec:tof}

We define a \textit{time-of-flow} as follows: Given a trace with a single, fixed \textit{root source signal}, the \textit{time-of-flow} of a \textit{sink signal} is Verilog time delta at which the \textit{tracking signal} of the \textit{sink signal} is set to a nonzero value.

We define a \textit{time-of-flow} tuple as the triple over the \textit{root source signal}, the \textit{sink signal}, and the Verilog time delta.

Considering only the \textit{root source signal} \texttt{key}, the \textit{times-of-flow} tuples of the design would be as follows:

\begin{center}
    
\begin{tabular}{llr}
\toprule
root & sink & $\Delta t$ \\
\midrule
key & one & 1 \\
key & two & 4 \\
key & out & 7 \\
\bottomrule
\end{tabular}
\end{center}

We note importantly that while these tuples can be regarded as (weighted) graph edges over the vertex set of signals, they do not necessarily correspond to atomic information flow paths. To return to the extended example, at time $\Delta t = 7$, when the \textit{tracking signal} \texttt{out\_t} is set to non-zero, this is not indicative of an information flow from the \textit{root source signal} \texttt{key} directly into \texttt{out}. Rather, the information flow path passed through the \textit{intermediate signal} \texttt{two} - though we do not yet have sufficient information to establish this path.

\subsubsection{\textit{Times-of-flow} Extraction}

While information flow paths are embedded within \textit{times-of-flow} from a single \textit{root source signal} as the transitive closure of all information flow from that fixed root. We wish to differentiate \texttt{sink signals} by \textit{intermediate signal} using the trace data we generated by simulating over all possible \texttt{root source signals}. We can imagine an extended set of a \textit{times-of-flow tuples}:

\begin{center}
    
\begin{tabular}{llr}
\toprule
root & sink & $\Delta t$ \\
\midrule
key & one & 1 \\
key & two & 4 \\
key & out & 7 \\
two & out & 7 \\
\bottomrule
\end{tabular}
\end{center}

There are no cases in which information flows from either \texttt{one} or \texttt{out} to any other signal, and only a single case in which any signal other than \texttt{key} flows - the single update at $\Delta t = 7$ when \texttt{out} is assigned the value of \texttt{two}. 

We show a subset of the tracking information in Figure~\ref{fig:all}. Here we show a counter module with different initial tracking states. Module \texttt{u1} is initialized with a \textit{root source signal} of \texttt{one}, and \texttt{u2} with \texttt{two} respectively. The otherwise identical modules are then executed using the same testbench. We see the anticipated results - there are no \textit{times-of-flow} generated from when exploring the signal \texttt{one} (a simple enough matter that could be concluded from static analysis, but can also be determined from the trace), and the single information leak of interest occurs at time $\Delta t = 7$ when considering signal \texttt{two}.

\subsubsection{Path Extraction}

Given the set of \textit{times-of-flow tuples} we can iteratively populate a graph of information flow relations between various registers. We imagine this as an iterative falsification process over an initial complete graph.

\underline{Candidate Edges:} We begin with the following, where dashed lines denote possible edges that have not yet been falsified, which we can perhaps regard as edges with an edge weight of $\varepsilon > 0$:
\begin{center}
\begin{tikzpicture}[
    node distance=2cm,
    every node/.style={circle, draw, minimum size=1cm, solid},
    every path/.style={->,dashed}
]

    \node (key) at (0,2) {key};
    \node (one) at (2,2) {one};
    \node (two) at (0,0) {two};
    \node (out) at (2,0) {out};

    \path (key) edge (one);
    \path (key) edge (two);
    \path (key) edge (out);

    \path (one) edge (key);
    \path (one) edge (two);
    \path (one) edge (out);

    \path (two) edge (key);
    \path (two) edge (one);
    \path (two) edge (out);

    \path (out) edge (key);
    \path (out) edge (one);
    \path (out) edge (two);

\end{tikzpicture}
\end{center}

\underline{Known Non-Edges:} We consider the \textit{times-of-flow tuples} and note that there are none such that the \textit{root source signal} is either \texttt{one} or \texttt{key}, so we omit all such as edges from the graph.

That is, given information flow graph $G = (V, E)$, where $V$ is the set of signals and $E$ is a set of pairs of signals representing information flow relations:
$$
\forall e = (v_{src}, v_{sink}) \in E, v_{src} \not\in \{\texttt{key}, \texttt{one}\} 
$$

We can apply a similar analysis to edges directed edges leaving the $v_{two}$, for which we know there is exactly one edge incident on $v_{out}$, so we may omit the others. We update our graph:

\begin{center}
\begin{tikzpicture}[
    node distance=2cm,
    every node/.style={circle, draw, minimum size=1cm, solid},
    every path/.style={->,dashed}
]

    \node (key) at (0,2) {key};
    \node (one) at (2,2) {one};
    \node (two) at (0,0) {two};
    \node (out) at (2,0) {out};

    \path (key) edge (one);
    \path (key) edge (two);
    \path (key) edge (out);

    \path (two) edge (out);

\end{tikzpicture}
\end{center}

\underline{Known Edges:} We can then consider the cases of vertices which correspond to non-empty sets of \textit{times-of-flow} tuples. We recognize that the tuple with the earliest time necessarily constitutes an atomic flow as there are not yet any \textit{intermediate signals}, so we consider for each \textit{root source signal} the set of known atomic flows $A$ over the set of \textit{times-of-flow tuples} $F$:

$$
A = \{ f = (v_0, v_1, \Delta t ) \in F : \forall f' = (v'_0, v'_1, \Delta t' ), \Delta t \leq \Delta t`\}
$$

This is perhaps simpler to view in the table of tuples:

\begin{center}
\begin{tabular}{lllr}
\toprule
index & root & sink & $\Delta t$ \\
\midrule
0 & key & one & 1 \\
1 & key & two & 4 \\
2 & key & out & 7 \\
3 & two & out & 7 \\
\bottomrule
\end{tabular}
\end{center}

Indices 0 and 3 have the minimal $\Delta t$ for their root, and therefore constitute singleton sets $A_{key}$ and $A_{two}$ respectively. With these as known atomic flows, we update our graph to denote that edges in these sets are known cases of atomic information flow.

\begin{center}
\begin{tikzpicture}[
    node distance=2cm,
    every node/.style={circle, draw, minimum size=1cm, solid},
    every path/.style={->}
]

    \node (key) at (0,2) {key};
    \node (one) at (2,2) {one};
    \node (two) at (0,0) {two};
    \node (out) at (2,0) {out};

    \path (key) edge (one);
    \path[dashed] (key) edge (two);
    \path[dashed] (key) edge (out);

    \path (two) edge (out);

\end{tikzpicture}
\end{center}

\underline{Candidate Promotion:} From there, we promote candidates to known flows in the case of an edge on some \textit{sink signal} for which a single \textit{source signal} is possible. In our example there is a single such case. It is known that at time $\Delta t = 4$ the signal \texttt{two} is a sink in the case where the \textit{root source signal} is \texttt{key}. Separately, it is known that the only possible source of information is \texttt{key}, as there is no possible path from any other tracked signal.

By regarding candidate flows as edges with weight $\varepsilon > 0$ and known atomic flows as edges with weight $1$, promotion is as simple as the following algorithm:

\begin{algorithmic}
\While{$\exists v \in V : \forall e = (v, v', x) \in E \implies x < 1$} 
    \If {$| \{ e = (v, v_1, x) \in E : x = \varepsilon\}| = 1$}
        \If{$\exists e = (v'', v, 1) \in E$}
            \State{$x \gets 1$}
        \EndIf
    \EndIf
\EndWhile
\end{algorithmic}

That is, while there are vertices with a single incoming candidate edge from a vertex where itself has at least one known incoming edge, regard said candidate edge as a known flow case.

\begin{center}
\begin{tikzpicture}[
    node distance=2cm,
    every node/.style={circle, draw, minimum size=1cm, solid},
    every path/.style={->}
]

    \node (key) at (0,2) {key};
    \node (one) at (2,2) {one};
    \node (two) at (0,0) {two};
    \node (out) at (2,0) {out};

    \path (key) edge (one);
    \path (key) edge (two);
    \path[dashed] (key) edge (out);

    \path (two) edge (out);

\end{tikzpicture}
\end{center}

\underline{Pathfinding Loopback:} Given a graph of known and candidate edges, we pass back over the \textit{time-of-flow tuples} for each root source signal. For each, we identify paths beginning with the root source signal for which each successive signal is (1) temporally later within the time of flow tuple and (2) along known edges from the prior signal in the path. These are all the known information flow paths over the trace.

\underline{Limitations:} We note that our example has one remaining candidate edge $(v_{key}, v_{out}, \varepsilon)$ that has not been removed through this method. In our case, there is no way to decisively conclude from trace data that there is in fact no flow in this case. 
Imagine a design with a more severe vulnerability containing an assignment such as the following:

\begin{verbatim}
        out <= two + key;
\end{verbatim}

In this case, we would add instrumentation as follows:

\begin{verbatim}
        out <= two + key;
        out_t <= two_t | key_t | out_t;
\end{verbatim}

And from the perspective of traces, the resultant value of \texttt{out\_t} would be identical. Remaining candidate edges document this limitation.

\section{Implementation}

For our implementation, we targeted open source Verilog designs for information flow path extraction. While we do not require verilog source access, we do minimal require (1) traces (2) specified in RTL (3) with IFT instrumentation, which for all practical purposes required the use of open-source designs.

\subsection{Information Flow Tracking}

Building on earlier work in automatic generation of information flow properties~\cite{deutschbein21, deutschbein2022, deutschbein2023} we use the \href{https://cycuity.com/type/fact-sheet/radix-automated-security-verification/}{Cycuity (née Tortuga Logic)'s Radix technology} which is based on prior work for RTL and gate-level IFT~\cite{ardeshiricham17,becker2017,hu2014,hu2016,hu2018}.

\begin{displayquote}
Cycuity’s Radix technology adds systematic hardware vulnerability detection and prevention to existing ASIC, SoC, and FPGA verification methodologies using its comprehensive information flow analysis technology.
\end{displayquote}

\subsection{Simulation}

\subsubsection{Testbench}

Testbench generation is a separate and active area of research~\cite{zheng24}, but we only require that some imperatives be dispatched to a hardware design from a simulation framework that may log the hardware state. Specifically, we adapt this requirement from the Myrtha package~\cite{deutschbein25hwcicd} which we use to parallelize simulation.

In our experience, the open source hardware designs we have studied, which include the AKER access control model~\cite{restuccia21}, and RISC-V designs such as \href{https://github.com/YosysHQ/picorv32}{PicoRV32} and \href{https://github.com/YosysHQ/nerv}{NERV} either came with a provided testbench in the design repository, or designers helpfully provided an example testbench for research purposes\footnote{Special thanks to Andres Meza.}.

\subsubsection{Setting the Source}

Radix uses a simple specification logic to programmatically track information flow. In additional to a hardware design and a testbench, we additionally furnish Radix with a information flow property or properties. In all cases, we used properties of the form:

\begin{verbatim}
autogenerated_<signal_name>: assert iflow (
    <signal_name>
    =/=>
    $all_outputs
);
\end{verbatim}

Where \texttt{<signal\_name>} was the name of some signal. In our example, these would be \texttt{key}, \texttt{out}, etc. and in the case of a CPU design would be the likes of \texttt{alu\_out} or \texttt{reg\_pc}.

We simulated the design over the provided testbench, extracted all registers (not all signals - we omitted Verilog \texttt{wire} signals), from the trace, and generated a property for each register. We then simulated the design, once each setting every unique register as the sole \textit{root source signal}. 

As a note, we importantly specified to Radix to check for information flow to all possible \textit{sink signals}. As an optimizing security tool, Radix would otherwise prune tracking to optimize performance unless prompting to considering all possible signals. Both the logical \texttt{=/=>} "does not flow" operator and the \texttt{\$all\_outputs} alias are language constructs specific to Radix, but could be translated to other tools.

This is logically equivalent to our manual example where we added an assignment to \textit{tracking signal} to be initially nonzero:

\begin{verbatim}
    key_t <= 1;
\end{verbatim}

\subsubsection{Trace Data}

We extract traces as IEEE 1364 VCD (value change dump) files, which can be used to determine the value of any register at any time point. It is these files we show in our figures using waveform visualizers such as \href{https://gtkwave.sourceforge.net/}{GTKWave} or \href{https://github.com/wavedrom/vcdrom}{VCDROM}.

Electing to use VCD files rather than LXT (interLaced eXtensible Trace) or the increasingly popular FST (Fast Signal Trace), which are also open standards, or the proprietary FSDB (Fast Signal DataBase, from Synopsys) or WLF (Wave Log Format, from Mentor), imposes a significant performance bottleneck.  That said, the VCD file format stores signal values as an ASCII binary representation, which is described by authors of the FST format as
\href{https://blog.timhutt.co.uk/fst_spec/}{"an ancient and hilariously inefficient text based format."}. Unfortunately, LXT is being deprecated; FST has no official specification; and the other formats are proprietary. So we used VCD.






\subsubsection{Mapping Stage}

For designs of any reasonable size, simulation for each signal within the design can be prohibitively expensive to both compute and store. We address compute costs through parallelization and storage costs through usage of binary storage formats.

Helpfully, due to the popularity and openness of the VCD format, we were able to architect around a costly storage stage for a text-based format. We used the vcd2df~\cite{deutschbein2025vcd2df} Python library to convert to binary (Parquet) format suitable for parallel computation. While not necessary for a minimized design, we elected for scalable technologies to target larger architectures.

First, we use the ``Myrtha"~\cite{deutschbein25hwcicd} package for automated
specification generation to enable cloud-scale parallelism of the trace generation process. In order for our method to work, we must generate a trace for every signal in the design. Modern designs have up to hundreds of billions of transistors, but parallelism makes the study of non-trivial designs feasible.

Second, we use the ``vcd2df"~\cite{deutschbein2025vcd2df} library in Spark
to compile the trace data into a single, cloud-native dataframe for analysis.
This dataframe collectively contains all \textit{time-of-flow} tuples across the entire design, regardless of \textit{root source signal}.
From there, we can query the dataframe to construct various graphs, including
candidate and known paths between pairs of signals. Critically, before performing any store operation, we may discard all information regarding the execution of the design, and store at most a single bit value per register, or use a sparse representation if doing so is more efficient.

\underline{Sparse Representation:}
In practice, we found the even binary representation was prohibitively large for storage and reference, so we developed a straightforward sparse representation using Python, which could easily be stored in Pickle files or as JSON.

We created a single object of the following type:
\begin{verbatim}
dict[str, dict[int, list[str]]]
\end{verbatim}

The outer dictionary had keys which were \textit{root source signals} and values which were inner dictionaries. The inner dictionaries had keys were time deltas ($\Delta t$) and values were lists of \textit{sink signals} . In this case, the special atomic set $A$, from our graph construction, was simply the inner dictionary entry with the minimal integer key.

We provide these scripts open-source on \href{https://github.com/cd-public/intcec26}{GitHub}.

\subsection{Graph Construction}

From the sparse representation, we reconstructed the graph as a matrix representation using a two-dimensional array of bits, where we used 0, 1 and -1 to denote non-edges, known edges, and candidates respectively.

To begin, we translated all trace data from VCD files to Python pickle files of dataframes. We performed this action in single Python script of 10 lines-of-code (LoC), essentially mapping the vcd2df package over the vcd directory.

\subsubsection{\textit{Times-of-flow} extraction}

We performed time-of-flow extraction with two filters (to remove non-IFT signals and signals always equal to zero) and a single query (idxmax, to find the first non-zero tracking value) over the dataframe of the trace data, then sorted the results by time-of-flow. We performed this action in single Python script of 18 LoC.

\subsubsection{Path Extraction}

We performed the entire path extraction in a single scripting stage to avoid cacheing a local result as a Python object rather than a dataframe. So we combined these algorithmic steps:
\begin{itemize}
    \item Known Non-Edges
    \item Known Edges
    \item Candidate Promotion
\end{itemize}

Separately, this was our first non-parallelizable step, beginning by looping over all the \textit{time-of-flow} dataframes. Fortunately, these dataframes were dramatically smaller, and with higher access speeds, than the VCD data.

We imagined this as a reduce operation (to form a single sparse object) and a construction operation (to build the graph. The reduce operation was implemented in 31 LoC, and then the construction stages were implemented as follows:

\begin{center}
\begin{tabular}{ c | c }
 Stage & LoC \\
 \hline
 Known Edges & 6 \\  
 Known Non-Edges & 6 \\ 
 Candidate Promotion & 7 \\
\end{tabular}
\end{center}

In total, graph construction from \textit{time-of-flow} tuples was implemented in 69 LoC including a few additional whitespace and linking lines, and contains all known and potential information flow paths over the design.

\underline{Pathfinding Loopback:} As we saved \textit{time-of-flow tuples} either within working memory or within the file system, it was a simple enough matter to loopback over the tuples and valid path subsets. We implemented what essentially amounted to an iterative path validator - where each adjacent pair of \textit{intermediate signals} where tested against the known edges - in 29 LoC, with a dependency on both graph construction \textit{times-of-flow} extraction. At this point, we saved our output as JSON (as enumerated the paths led to unwieldy files, as JSON has strong software support).

\section{Case Study}

We applied our technique to the PicoRV32 RISC-V CPU. As we used a proprietary IFT technology (Radix), we used open-source VCD files from the Isadora specification miner~\cite{deutschbein2022}, available publicly on \href{https://github.com/cd-public/intcec26}{GitHub} for reproducibility. PicoRV32 contains 232 registers in 3049 lines of Verilog code. Its accompanying testbench is 86 lines of Verilog and runs for 2201 cycles. Of the 232 registers, 182 have nontrivial information flow profiles (at least one outgoing flow).

\subsection{Mapping Stage}

We generated traces serially for all 182 nontrivial registers as \textit{root source signal} in 8h45m. We did not parallelize as we lacked access to the appropriate license for our IFT simulator to do so. However, with each register taking around 3 minutes (average of 2:53) to simulate, and the Myrtha~\cite{deutschbein25hwcicd} package average of 34.1 to initialize time on cloud runners, we would expect a comically parallel end-to-end trace generation time of around 3:27 seconds on academic-tier Azure/GitHub Actions.

IFT instrumented trace files are quite memory intensive, with the total storage requirements for the VCD files totaling 64 MB and their dataframes coming in at 525 MB. Of course, we do not need to cache dataframes and did so in this case only for reproducibility - an individual dataframe would trivially fit in the cache of a runner, and could be converted to a sparse representation before the reduce operation.

Serialized dataframe generation took 28.9 seconds for the full design, or .16 seconds per register. This was an extremely low variance operation - the VCD files never differed by more than a half percent in size by \textit{root source signal}, and translation time was dominated by text parsing. Ultimately, usage of dataframes here would have a barely noticeable impact compared to the much more costly trace generation, but made analysis of the trace data much easier.

\subsection{Sparse Representation}

The spare representation was a straightforward but unexpectedly performant implementation for PicoRV32. Of the 182 sparse representations of \textit{times-of-flow tuples} per \textit{root source signal}, not a single file was larger than 6 KB, the average was 2.3 KB, and total storage space was only 419 KB. This represented a space savings of over 1000x versus full dataframes, and 100x vs VCD files.

Serialized sparse encoding took 1.531 seconds for the whole design. Individual registers took around .3 seconds to encode because per-register time was so low and loading our dataframe library (pandas) was relatively heavyweight. That said, the sparse encoding could be parallelized along with the VCD read step, and would add only 7ms in compute time, plus the time savings from writing a 1000x smaller pickle file. It was difficult to extrapolate performance difference this narrow over a smaller open-source design. However, we found bypassing the intermediate dataframe stage and writing sparse representations directly took the same amount of time to three significant figures (vs. using the intermediate dataframe during serialized testing). Comparatively costly sparse compute operations were offset by the less costly access operations.

As an implementation detail, we shifted to using an array rather than JSON or built-in Python data storage types for interoperability with the dataframe format.

\subsection{Graph Construction}

We trivially store the full graph as an adjacency matrix with byte representation (we used 0 for no edge, 1 for known edge, -1 for candidate edge). This resulted in a single file of 36 KB, where we expect 33 KB to have been the adjacency matrix and the remaining 3 KB encoding the labels (signal names). We vectorized this operation for the following time cost:
\begin{verbatim}
$ time python3 sparse_reduce.py
real    0m0.846s
user    0m2.231s
sys     0m0.135s
\end{verbatim}

\underline{Pathfinding Loopback:} We stored the paths as trees, rather than individual paths, requiring only 32 KB to store all paths in ASCII encoded JSON. Coincidentally, JSON encoding and the adjacency matrix had the same storage cost for this design, which we hope to explore on different types and sizes of designs in future work to better assess scalability. This stage could be partially parallelized though we performed it serially after constructing the adjacency matrix, and found the performance appropriate while serialized for our design size.

\begin{verbatim}
$ time python3 paths.py
real    0m0.445s
user    0m2.192s
sys     0m0.036s
\end{verbatim}

\subsection{Results}

In total, we found known paths of various lengths as follows:

\begin{center}
\begin{tabular}{lrrrr}
\toprule
length & 2 & 3 & 4 & 5 \\
\midrule
count & 151 & 130 & 71 & 59 \\
\bottomrule
\end{tabular}
\end{center}

As an example, we found the following path:

\begin{verbatim}
do_waitirq
  ==> mem_do_rinst 
  ==> mem_valid 
  ==> next_insn_opcode
  ==> dbg_insn_opcode
\end{verbatim}

This shows pending interrupt requests are observable via debug ports - which, if unintended, is \href{https://cwe.mitre.org/data/definitions/1244.html}{CWE-1244: Internal Asset Exposed to Unsafe Debug Access Level or State}.

\section{Extended Case Study}

A partial motivation to develop path reconstruction over trace data was the limitation of the earlier Isadora tool~\cite{deutschbein2022}. In this prior work, validation was able to find security properties which were consistent with, but not necessarily evidence of, a confused deputy. In our work, now, with the benefit of path construction, we are able to derive a precise path and show a violation of  CWE 411: Unintended Proxy or Intermediary (‘Confused Deputy’).

We studied a design with two access control modules and three peripherals implementing this access control policy:

\begin{align*}
\text{AC}_1 \text{ of ACW}_1 & : \quad R = \{P_1, P_2\}, \, W = \{P_1\} \\
\text{AC}_2 \text{ of ACW}_2 & : \quad R = \{P_3\}, \, W = \{P_2, P_3\}
\end{align*}

Using our new tool, we are able to find an atomic path between registers of $\text{ACW}_2$ and registers of $P_1$. This is a violation of the access control policy which allows no write access between these entities. Prior work should that information flowed from $\text{ACW}_2$ to $P_1$ but lacked the expressive power to show it flowed directly (that is, could have flowed along some legal path using intermediaries) and therefore could not decisively establish whether a security violation occurred. However, an atomic path, which we discover, captures what is necessarily security violation.

\section{Discussion}

This algorithmic approach to graph reconstruction is specific to working on trace data containing IFT signals, which leads to a number of strengths and weaknesses.

\subsection{Trace Dependence}

Traces can adversely impact the usefulness of discovered traces if test coverage is poor, such as of insufficiently scope or because it is unrepresentative and misses important cases. This is an inherent weakness of any trace-based methodology, but also poses some minor upside.

By studying traces, this technique necessarily is restricted to information flows which actually occur - that is, there are no theoretical flows which do not occur in practice. As a component of a broader validation effort, trace-based analysis can therefore provide a useful refinement to other techniques.

In prior work, this formulation has been used to view analysis as not being restricted just to hardware, but both hardware and its operating system together~\cite{deutschbein20}. 

Lastly, we note that while our work uses a testbench~\cite{meng22} for trace data, testbench generation is an
active area of research, and is more fully explored in related works
such as Meng et al. [33], which studies concolic testing for RTL.

\subsection{Competing Technologies}

For other techniques to find information flow paths, the two most common we find are static analysis~\cite{deutschmann25} and symbolic execution~\cite{ryan23fmcad, ryan23seif}. Both are usually paired with formal methods.

Static analysis helpfully finds all possible information flow path, but often finds a superset thereof as not all design states are practically reachable within most modern designs. In practice, we find the majority of possible hardware states to result an immediate crash, yet these states would be included within naive static analysis. While recent techniques have begun to address these limitations, the use of traces is itself one way to find what necessarily occurs in practice.

Symbolic execution splits the difference between traces and static analysis by considering the universe of possible hardware states from some valid initial stage, and therefore most closely captures the set of all hardware states reachable in practice. However, symbolically executing hardware can be prohibitively computational expensive. While recent techniques have begun to address these limitations, the use of concrete traces is one way to manage this complexity.

\subsection{Understanding Candidate Edges}

We use the term candidate edges from the perspective of a graph algorithm, but in the context of security validation efforts they can be understood more properly as the set of signals which could be joined by logical disjunction (inclusive or) as in our earlier example:

\begin{verbatim}
      two   <= key - one;
      two_t <= key_t | one_t | two_t;
\end{verbatim}

Because it is so often the case that information flows from multiple signals simultaneously, these candidate edges will necessarily exist in practice. We expect the most common way to understand these edges in context is by exploring the design further with some other technology, such as static analysis or symbolic execution. It is also possible in some case to generate of a testbench that drives control flow down alternative paths from the initial exploration to perhaps have only one of the possible signals have a non-zero tracking label. We consider these promising directions for collaboration and future work.

\subsection{Parallelization}

It is simplest to understand performance gains to parallelization at the trace generation stage, which is by far the most expensive stage at this time. For PicoRV32, trace generation took 8h35m. When parallelizing all
trace generation, path construction could theoretically
evaluate the PicoRV32 design fully in less than five minutes. 
Each trace is generated in approximately 100 seconds, but are fully parallelizable. 
Further, the trace generation time is dominated
by write-to-disk, and performance engineering techniques could
likely reduce it significantly, such as by changing trace encoding
or piping directly to later phases, most likely the spare representation. This is roughly a 100x speedup for small designs, and should provide greater benefits on large designs if it is practical to provision enough instances of the simulator.

In future work we intend to modify the simulation framework to stream trace events, rather than write-to-disk, and perform validation online during simulation to remove the performance bottleneck in our end-to-end process.

\section{Conclusion}

We demonstrate a technique for extracting information flow paths from trace data on open-source hardware designs, we can be used to explore a wide range of hardware vulnerabilities. We demonstrate the ability to recognize CWEs for a design provided with design source and a testbench.





\section{Acknowledgements}
We thank the NSF for funding. We thank Cycuity, the Kastner Research Group of UCSD and HWSec@UNC research group for coordinating access to the Radix IFT technology.

\bibliographystyle{unsrt}
\bibliography{refs}

\end{document}